\documentclass[final, 5p, times, twocolumn]{elsarticle}

\usepackage{amsmath,amsfonts,amssymb,bm}
\usepackage{color}
\usepackage{dsfont}
\usepackage{graphicx}
\usepackage{multirow}
\usepackage{soul}

\usepackage[mathscr,scaled=1.15]{urwchancal}
\DeclareFontFamily{OT1}{pzc}{}
\DeclareFontShape{OT1}{pzc}{m}{it}%
{<-> s * [1.15] pzcmi7t}{}
\DeclareMathAlphabet{\mathpzc}{OT1}{pzc}{m}{it}

\definecolor{purple}{rgb}{0.5,0,0.5}
\definecolor{blue}{rgb}{0.0,0,0.9}
\definecolor{prdblue}{rgb}{0.133,0.118,0.498}
\usepackage[colorlinks=true, pdfstartview=FitV, linkcolor=prdblue, citecolor= prdblue, urlcolor=prdblue]{hyperref}

\biboptions{sort&compress}

\journal{Physics Letters B}

\begin{document}

\begin{frontmatter}



\title{$\,$\\[-7ex]\hspace*{\fill}{\normalsize{\sf\emph{Preprint no}. NJU-INP 027/20}}\\[1ex]
Form Factors of the Nucleon Axial Current}


\author[GU]{Chen Chen}
\ead{Chen.Chen@theo.physik.uni-giessen.de}

\author[GU,HFHF]{Christian S. Fischer}
\ead{christian.fischer@theo.physik.uni-giessen.de}

\author[NJU,INP]{Craig D. Roberts\corref{cor2}}
\ead{cdroberts@nju.edu.cn}

\author[UPO,INP]{Jorge Segovia}
\ead{jsegovia@upo.es}

\address[GU]{
Institut f\"ur Theoretische Physik, Justus-Liebig-Universit\"at  Gie\ss en, D-35392 Gie\ss en, Germany}

\address[HFHF]{
Helmholtz Forschungsakademie Hessen f\"ur FAIR (HFHF),
    GSI Helmholtzzentrum f\"ur Schwerionenforschung, Campus Gie{\ss}en,
    35392 Gie{\ss}en, Germany}

\address[NJU]{
School of Physics, Nanjing University, Nanjing, Jiangsu 210093, China}
\address[INP]{
Institute for Nonperturbative Physics, Nanjing University, Nanjing, Jiangsu 210093, China}

\address[UPO]{Dpto. Sistemas F\'isicos, Qu\'imicos y Naturales, Univ.\ Pablo de Olavide, E-41013 Sevilla, Spain}

\begin{abstract}
A symmetry-preserving Poincar\'e-covariant quark+diquark Faddeev equation treatment of the nucleon is used to deliver parameter-free predictions for the nucleon's axial and induced pseudoscalar form factors, $G_A$ and $G_P$, respectively.
The result for $G_A$ can reliably be represented by a dipole form factor characterised by an axial charge $g_A=G_A(0)=1.25(3)$ and a mass-scale $M_A = 1.23(3) m_N$, where $m_N$ is the nucleon mass; and regarding $G_P$, the induced pseudoscalar charge $g_p^\ast = 8.80(23)$, the ratio $g_p^\ast/g_A = 7.04(22)$, and the pion pole dominance \emph{Ansatz} is found to provide a reliable estimate of the directly computed result.
The ratio of flavour-separated quark axial charges is also calculated: $g_A^d/g_A^u=-0.16(2)$.  This value expresses a marked suppression of the size of the $d$-quark component relative to that found in nonrelativistic quark models and owes to the presence of strong diquark correlations in the nucleon Faddeev wave function -- both scalar and axial-vector, with the scalar diquark being dominant.
The predicted form for $G_A$ should provide a sound foundation for analyses of the neutrino-nucleus and antineutrino-nucleus cross-sections that are relevant to modern accelerator neutrino experiments.
\\[1ex]
\leftline{2021 February 15}
\end{abstract}




\end{frontmatter}


\section{Introduction}\label{SecIntro}
\unskip
In a step beyond the Standard Model, it is now known that neutrinos have mass \cite{RevModPhys.88.030501, RevModPhys.88.030502}; and it may be that this mass is not generated by couplings to the Higgs boson \cite{Gouvea:2016shl}.  The mass splittings between different neutrino species and the angles that describe mixing between them have been measured with precision \cite[Sec.\,14]{Zyla:2020}.  Yet, there are unresolved problems, \emph{inter alia}: what are the masses of the individual neutrinos; are neutrinos their own antiparticles; and have neutrinos played a pivotal role in Universe evolution?  With these questions bearing upon fundamental mysteries in Nature, high profile experiments are underway, being constructed or planned worldwide \cite[Sec.\,14]{Zyla:2020}.

The analysis and reliable interpretation of modern neutrino experiments relies on sound theoretical knowledge of neutrino/antineutrino-nucleus ($\nu/\bar \nu$-$A$) interactions \cite{Mosel:2016cwa, Alvarez-Ruso:2017oui, Hill:2017wgb, Gysbers:2019uyb, Lovato:2020kba, King:2020wmp}.  An important element in such calculations is the nucleon axial form factor, $G_A(Q^2)$, whose value at $Q^2=0$ is the nucleon's nonsinglet axial charge, $g_A=1.2756(13)$ \cite{Zyla:2020}, which determines the rate of neutron-to-proton $\beta$-decay: $n\to p + e^- + \bar\nu$.  Significantly, at the structural level, $g_A$ measures the difference between the light-front number-density of quarks with helicity parallel to that of the nucleon and the density of quarks with helicity antiparallel \cite{Deur:2018roz}.

$G_A(Q^2)$ has long been a subject of interest.  It was extracted from $\nu p$ and $\nu$-deuteron, $d$, scattering experiments performed over thirty years ago \cite{Baker:1981su, Miller:1982qi, Kitagaki:1983px, Ahrens:1986xe}, yielding results that are consistent with dipole behaviour characterised by a mass-scale $M_A \approx 1.1\,m_N$, where $m_N$ is the nucleon mass.  This value was also obtained in a more recent $p(e,e^\prime \pi^+)n$ experiment \cite{Liesenfeld:1999mv} and in a new analysis of the world's data on $\nu d$ scattering, albeit with a larger uncertainty than estimated in the original analyses \cite{Meyer:2016oeg}.
On the other hand, modern experiments using $\nu$ scattering on an array of heavy targets (water, iron, mineral oil, Kevlar, and carbon) yield results covering the range $1.1 \lesssim M_A/m_N \lesssim 1.60$ \cite{Gran:2006jn, Dorman:2009zz, AguilarArevalo:2010zc, Fields:2013zhk, Fiorentini:2013ezn}.  Important issues here are the reliability of the model used to describe the nuclear target and differences between the models used by the collaborations.

It is conceivable that the properties of individual nucleons are modified when they are embedded in a nucleus.  That is one possible explanation of the data in Ref.\,\cite{Aubert:1983xm}.  On the other hand, evidence supporting an explanation via short range nucleon-nucleon correlations within a nucleus \cite{Weinstein:2010rt} suggests that if in-medium modifications do occur, then only a small fraction of the number of nucleons is affected.  Moreover, data on electron scattering from nuclei on a wide kinematic domain is readily explained by modern many-body methods in nuclear physics using in-vacuum nucleon properties, see \emph{e.g}.\ Ref.\,\cite{Lynn:2019rdt}.  Hence, accurate information on the single-nucleon axial form factor is a necessary element in a successful description of $\nu A$ scattering \cite{Gysbers:2019uyb, Lovato:2020kba, King:2020wmp}.  It may not be sufficient.

Nucleon and $\Delta$-baryon form factors have been studied using an array of tools \cite{DeSanctis:1998ck, DeSanctis:2011zz, Hagler:2009ni, Punjabi:2015bba}; and there have been many model analyses of the nucleon's axial current, \emph{e.g}.\ Refs.\,\cite{DeGrand:1975cf, Thomas:1981vc, Weise:1991, Alkofer:1994ph, Boffi:2001zb, Adamuscin:2007fk} and citations thereof.  Today, numerical simulations of lattice-regularised quantum chromodynamics (QCD) are also being deployed to determine $G_A(Q^2)$.  The contemporary status of such analyses is summarised in Ref.\,\cite{Kronfeld:2019nfb}: the results correspond to dipole masses in the range $1.1\lesssim M_A/m_N \lesssim 1.7$.

Evidently, considering both experiment and theory, $M_A$ is not known to better than 50\%.  It has been argued \cite{Hill:2017wgb} that if the precision can be increased to 10\% or better, then $G_A$ will become a subdominant source of error in the determination of neutrino properties in $\nu$-oscillation experiments.  This is good motivation for a new analysis of the nucleon's axial current form factors.

Continuum Schwinger function methods \cite{Eichmann:2016yit, Papavassiliou:2017qlq, Huber:2018ned, Fischer:2018sdj, Roberts:2020hiw, Qin:2020rad} have been used in an extensive analysis of the nucleon's axial current \cite{Eichmann:2011pv}; but being based on the leading-order truncation (rainbow-ladder, RL \cite{Munczek:1994zz, Bender:1996bb}) of the associated quantum field equations, important contributions tied to the emergence of hadronic mass (EHM) are underestimated \cite{Chang:2012cc}.  Therefore, we choose to approach the problem using the Poincar\'e-covariant quark+diquark Faddeev equation framework that has successfully been employed in the description and unification of an array of properties of the nucleon, $\Delta$-baryon, and their low-lying excitations, \emph{viz}.\ masses, wave functions, and elastic and transition form factors \cite{Segovia:2014aza, Burkert:2017djo, Chen:2017pse, Chen:2018nsg, Chen:2019fzn, Lu:2019bjs, Cui:2020rmu}.  With its inputs tuned elsewhere via comparisons with different observables, many aspects of EHM are implicitly expressed in the formulation.

\section{Nucleon Axial Current}
\label{SecNAC}
The nucleon's isovector axial-vector current is
\begin{subequations}
\label{EqJ5A}
\begin{align}
J^j_{5\mu}(K,&Q)  = \bar u(P_f) \tfrac{1}{2}\tau^j \Lambda_{5\mu}(K,Q) u(P_i) \\
& = \bar u(P_f)\gamma_5 \tfrac{1}{2}\tau^j \bigg[\gamma_\mu G_A(Q^2) +i \frac{ Q_\mu }{2 m_N} G_P(Q^2)\bigg] u(P_i) \,,
\end{align}
\end{subequations}
where we have assumed isospin symmetry,
$\{\tau^i|i=1,2,3\}$ are Pauli matrices,
$K=(P_f+P_i)/2$, $Q=(P_f-P_i)$,
$P_f^2=-m_N^2=P_i^2$,
$G_A(Q^2)$ is the axial form factor,
and $G_P(Q^2)$ is the induced pseudoscalar form factor.  The kindred pseudoscalar current is
\begin{subequations}
\label{EqJ5}
\begin{align}
J^j_{5}(K,Q)  & = \bar u(P_f) \tfrac{1}{2}\tau^j \Lambda_{5}(K,Q) u(P_i) \\
& = \bar u(P_f)\gamma_5 \tfrac{1}{2}\tau^j G_5(Q^2) u(P_i) \,.
\end{align}
\end{subequations}
This pair of currents is related by an axial-vector Ward-Green-Takahashi identity:
\begin{equation}
\label{EqAVWGTIa}
Q_\mu J^j_{5\mu}(K,Q) + 2 i m_q J^j_{5}(K,Q) = 0\,,
\end{equation}
where $m_q$ is the current-quark mass; and this entails
\begin{equation}
\label{WGTIexplicit}
2 m_N G_A(Q^2) - \frac{Q^2}{2 m_N} G_P(Q^2) = 2 m_q G_5(Q^2)\,.
\end{equation}
Eq.\,\eqref{WGTIexplicit} imposes tight constraints on any analysis of these form factors.

\begin{figure}[t]
\centerline{%
\includegraphics[clip, width=0.45\textwidth]{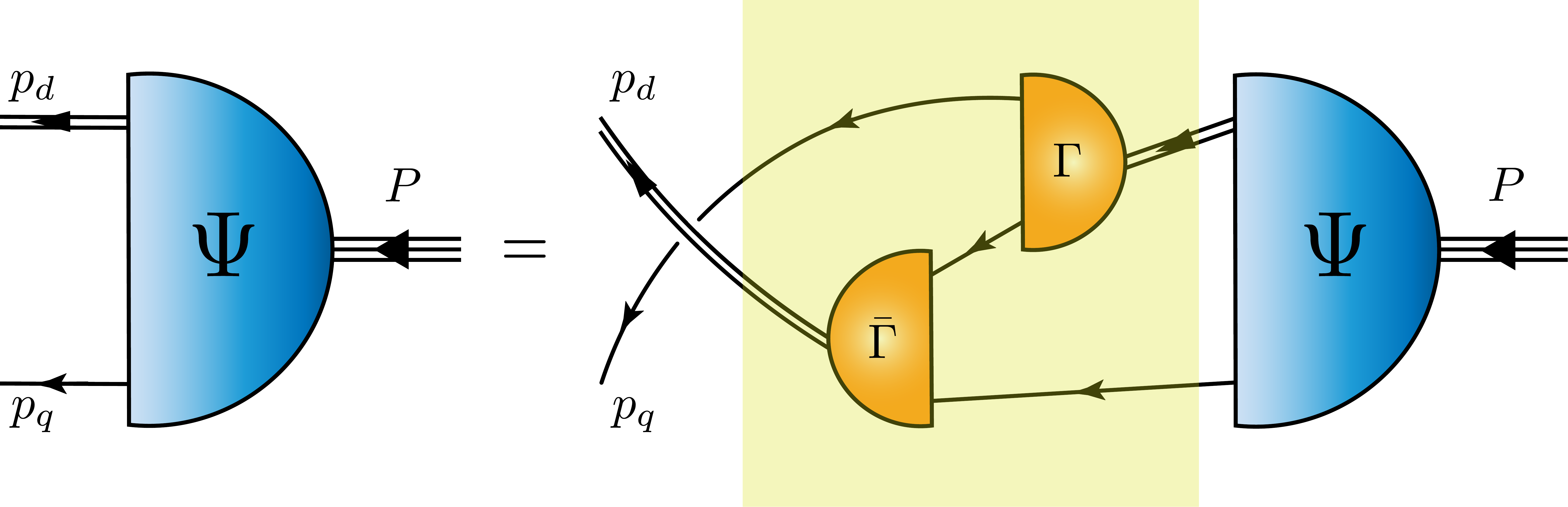}}
\caption{\label{FigFaddeev}
Quark+diquark Faddeev equation, a linear integral equation for the Poincar\'e-covariant matrix-valued function $\Psi$, the Faddeev amplitude for a nucleon with total momentum $P= p_q + p_d$.  $\Psi$ describes the relative momentum correlation between the dressed-quarks and -diquarks.  Legend. Shaded rectangle -- Faddeev kernel; \emph{single line} -- dressed-quark propagator; $\Gamma$ -- diquark correlation amplitude; and \emph{double line} -- diquark propagator.
Ground-state nucleons ($n$ - neutron, $p$ - proton) contain both isoscalar-scalar diquarks, $[ud]\in (n,p)$, and isovector--axial-vector diquarks $\{dd\}\in n$, $\{ud\}\in (n,p)$, $\{uu\}\in p$.
Other possible correlations play no significant role in nucleons \cite{Eichmann:2016yit, Barabanov:2020jvn}.
}
\end{figure}

\smallskip

\hspace*{-\parindent}\emph{Faddeev Amplitude}\,---\,%
The first step in the calculation of $G_{A,P}$ is computation of the nucleon's mass and Poincar\'e-covariant bound-state amplitude from the quark+diquark Faddeev equation introduced in Refs.\,\cite{Cahill:1988dx, Reinhardt:1989rw, Efimov:1990uz} and depicted in Fig.\,\ref{FigFaddeev}.  There is a large body of evidence supporting the presence of diquark correlations within baryons, based on experiment, phenomenology, and theory \cite{Barabanov:2020jvn}.
Importantly, the diquarks in Fig.\,\ref{FigFaddeev} are nonpointlike and fully dynamical: they appear in a Faddeev kernel -- shaded domain -- which requires their continual breakup and reformation, and participate in interactions with all probes as allowed by their quantum numbers.

In solving the Faddeev equation, we use the following diquark masses (in GeV):
\begin{equation}
\label{Eqqqmasses}
m_{[ud]} = 0.80\,,\;
m_{\{uu\}} = m_{\{ud\}} = m_{\{dd\}} = 0.89\,;
\end{equation}
and light-quarks characterised by a Euclidean constituent mass $M_{q=u,d}^E = 0.33\,$GeV.  The associated propagators and additional details concerning the Faddeev kernel are presented in Ref.\,\cite[App.\,1, App.\,2]{Segovia:2014aza}.  Importantly, \emph{e.g}.\ the light-quark mass function, illustrated in Ref.\,\cite[Fig.\,6]{Chen:2017pse}, is in fair agreement with that obtained in modern gap equation studies \cite{Chang:2010hb, Chang:2011ei, Williams:2015cvx}.

These inputs are sufficient to obtain the nucleon Faddeev amplitude, $\Psi$, and mass $m_N = 1.18\,$GeV.  This mass value is intentionally large because Fig.\,\ref{FigFaddeev} describes the nucleon's \emph{dressed-quark core}.  To obtain the complete nucleon, resonant contributions should be included in the Faddeev kernel.  Such ``meson cloud'' effects generate the physical nucleon, whose mass is roughly $0.2$\,GeV lower than that of the core \cite{Ishii:1998tw, Hecht:2002ej, Sanchis-Alepuz:2014wea}.  Their impact on nucleon structure can be incorporated using sophisticated dynamical coupled-channels models \cite{Aznauryan:2012ba, Burkert:2017djo}, but that is beyond the scope of contemporary Faddeev equation analyses.
Instead, we express all form factors in terms of $x= Q^2/m_N^2$, a procedure that has proved efficacious in developing sound comparisons with nucleon and nucleon-resonance electromagnetic observables \cite{Segovia:2014aza, Burkert:2017djo, Chen:2017pse, Chen:2018nsg, Chen:2019fzn, Lu:2019bjs, Cui:2020rmu}.  In addition, we report a model uncertainty obtained by independently varying the diquark masses by $\pm 5$\%, thereby changing $m_N$ by $\pm 3$\%.

\smallskip

\begin{figure}[t]
\centerline{%
\includegraphics[clip, width=0.45\textwidth]{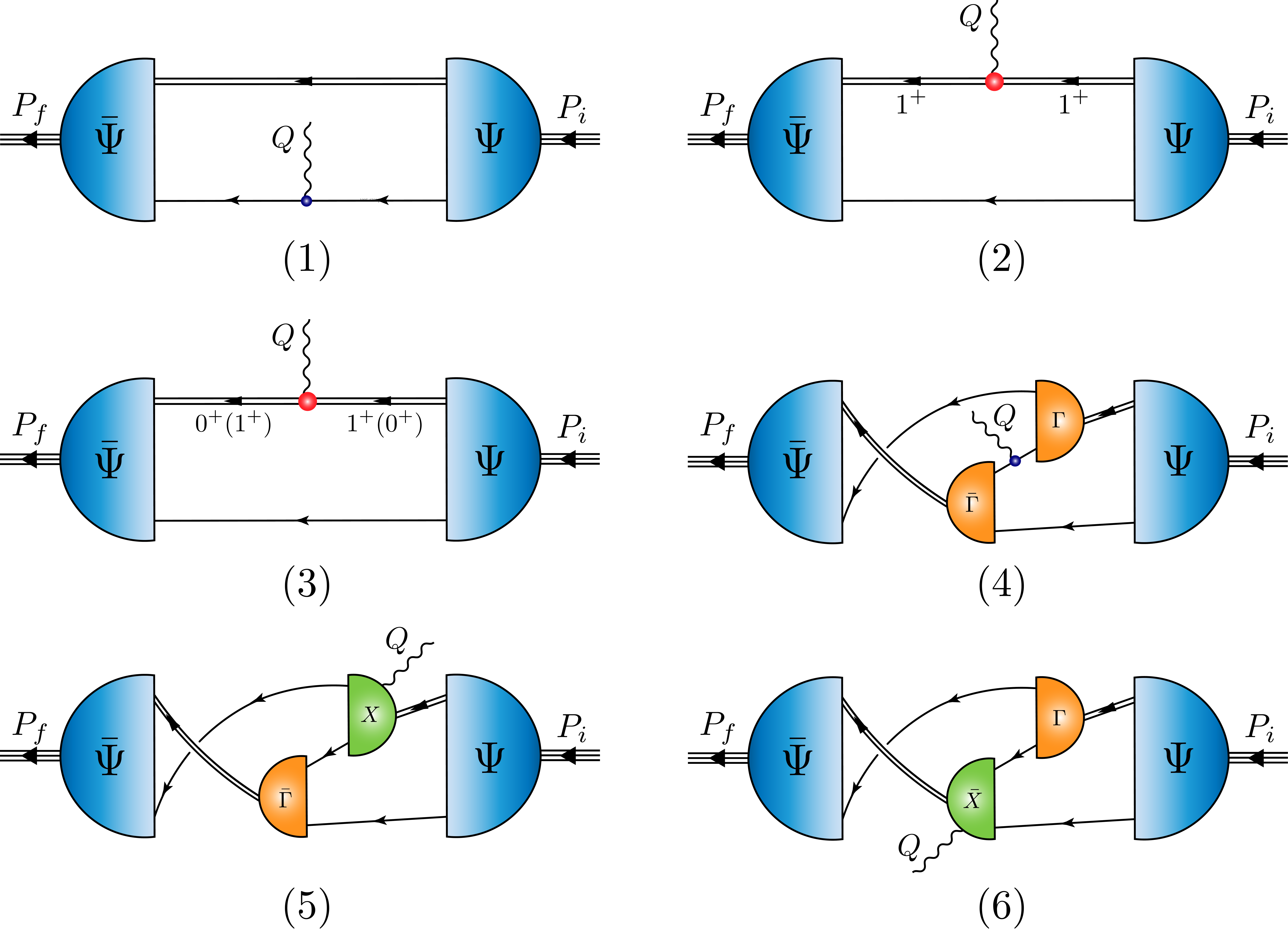}}
\caption{\label{FigCurrent}
Electroweak interaction vertex that ensures the on-shell nucleon obtained as the solution of Fig.\,\eqref{FigFaddeev} satisfies the axial-vector Ward-Green-Takahashi identity, Eq.\,\eqref{EqAVWGTIa}: \emph{single line}, dressed-quark propagator; \emph{undulating line}, weak boson; $\Gamma$,  diquark correlation amplitude; \emph{double line}, diquark propagator; and $\chi$, seagull interaction.
$\bar\Gamma(k,Q) = C^\dagger \Gamma(-k,Q)^{\rm T} C$, where $C$ is the charge conjugation matrix in spinor space and ``T'' indicates matrix transpose.  $\bar\chi$ is analogously determined by $\chi$.
%
%
(Details are provided in Ref.\,\cite{Chen:2020:progress})
}
\end{figure}

\hspace*{-\parindent}\emph{Current Elements}\,---\,%
The next step requires construction of the nucleon weak interaction current.  The minimal form consistent with the axial-vector Ward-Green-Takahashi identities at the quark, diquark, and nucleon levels can be derived following the procedures in Refs.\,\cite{Oettel:1999gc, Bloch:1999rm, Oettel:2000jj}.  It is depicted in Fig.\,\ref{FigCurrent} and explained in Ref.\,\cite{Chen:2020:progress}.  Herein, we recapitulate the salient details.

The vertex describing interactions between a weak-boson and dressed-quark in Fig.\,\ref{FigCurrent} -- diagrams\,(1) and (4) is \cite{Bloch:1999rm}:
\begin{subequations}
\label{EqWeakq}
\begin{align}
\Gamma_{5\mu}^j(k_+,k_-) &=  \gamma_5\frac{\tau^j}{2}\bigg[\Gamma_{5\mu}^{{\rm A}}(k_+,k_-) +
 \frac{2 i Q_\mu}{Q^2+m_\pi^2} \Sigma_B(k_+^2,k_-^2)\bigg] \,,\\
 \Gamma_{5\mu}^{{\rm A}}(k_+,k_-) & =
 \gamma_\mu \Sigma_A(k_+^2,k_-^2) + 2 k_\mu \gamma\cdot k \Delta_A(k_+^2,k_-^2)\,,
\end{align}
\end{subequations}
where $Q$ is the incoming momentum of the weak boson, $k_\pm = k\pm Q/2$, $m_\pi \approx 0.14\,$GeV is the pion mass, $\Sigma_F(k_+^2,k_-^2) = [F(k_+^2+F(k_-^2) ]/2$, $\Delta_F(k_+^2,k_-^2) = [F(k_+^2)-F(k_-^2) ]/[k_+^2-k_-^2]$, $F=A,B$.

Since the diquark correlations are nonpointlike and fully interacting, symmetry preservation requires that the nucleon electroweak current include seagull interactions, diagrams (5) and (6) in Fig.\,\ref{FigCurrent}.  Their explicit form can be derived by adapting the procedure in Ref.\,\cite{Oettel:1999gc}, which considered the analogous problem for the nucleon electromagnetic current.  With the Faddeev equation defined in Ref.\,\cite{Segovia:2014aza}, the electroweak seagulls in diagram (5) and (6) are, respectively \cite{Chen:2020:progress}:
\begin{subequations}
\label{EqSeagull}
\begin{align}
\chi_{5\mu}^j(k,Q) & = -\frac{iQ_\mu}{Q^2+m_\pi^2} \nonumber \\
& \times \bigg[
\gamma_5\frac{\tau^j}{2}\Gamma^{J^P}(k-Q/2)+
\Gamma^{J^P}(k+Q/2) \bigg(\gamma_5\frac{\tau^j}{2}\bigg)^{\rm T}\bigg], \\
\bar\chi_{5\mu}^j(k,Q) & = -\frac{iQ_\mu}{Q^2+m_\pi^2} \nonumber \\
& \times \bigg[\bar\Gamma^{J^P}(k+Q/2) \gamma_5\frac{\tau^j}{2}
+\bigg(\gamma_5\frac{\tau^j}{2}\bigg)^{\rm T} \bar \Gamma^{J^P}(k-Q/2)\bigg],
\end{align}
\end{subequations}
where $\Gamma^{J^P}$ is the correlation amplitude of the interacting diquark.
Since these terms are both proportional to $Q_\mu$, they do not contribute to $G_A$, which is determined by the $Q$-transverse piece of the nucleon axial current.

\begin{figure}[t]
\centerline{%
\includegraphics[clip, width=0.45\textwidth]{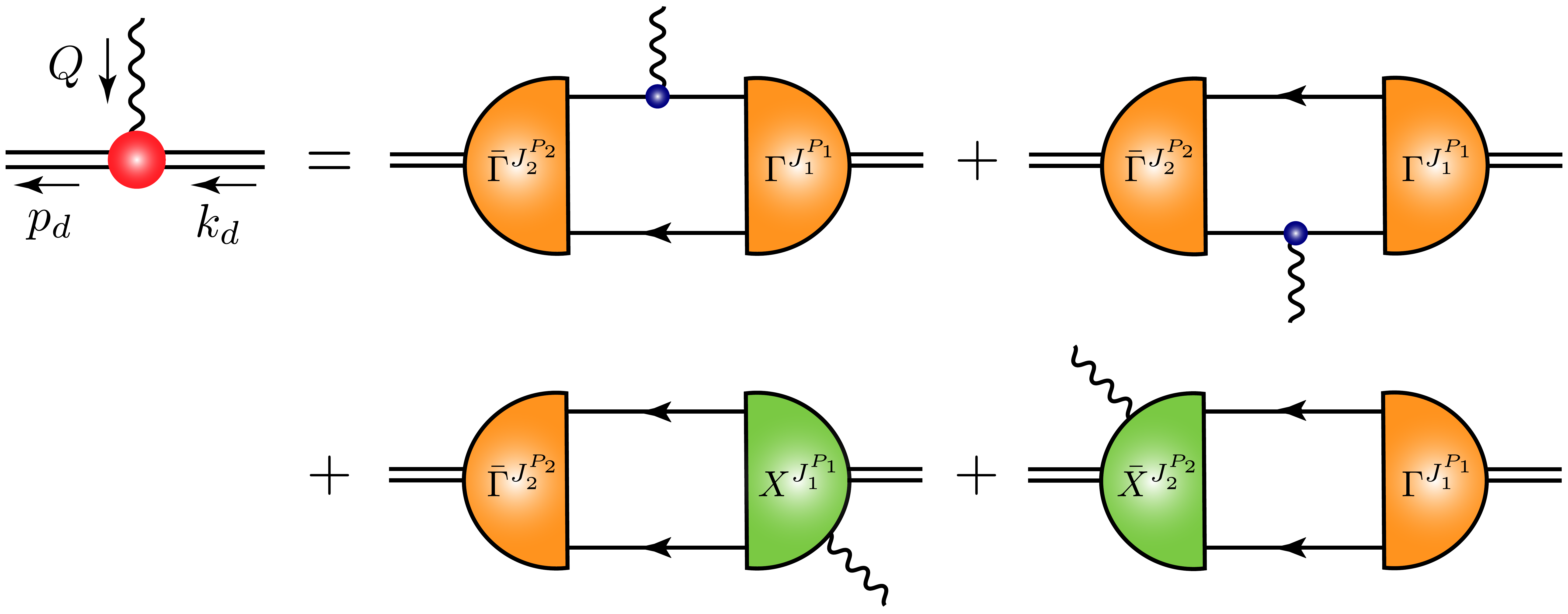}}
\caption{\label{FigWZqq}
Interaction vertex for the $J_1^{P_1} \to J_2^{P_2}$ diquark transition ($p_d=k_d+Q$): \emph{single line}, quark propagator; \emph{undulating line}, weak boson; $\Gamma$,  diquark correlation amplitude; \emph{double line}, diquark propagator; and $\chi$, seagull interaction.
}
\end{figure}

All that remains to complete the current is determination of the weak-boson coupling to diquarks, indicated by diagrams (2) and (3) in Fig.\,\ref{FigCurrent}.  These couplings are expressed by the sum of diagrams in Fig.\,\ref{FigWZqq}.
Given that the scalar diquark correlation is also isoscalar, associated with $\tau^2$, then there is no $0^+ \to 0^+$ contribution to the electroweak current.

On the other hand, transitions between scalar and axial-vector diquarks are possible: charged-currents -- $\{dd\} \leftrightarrow [ud]$, $[ud]\leftrightarrow \{uu\}$; and neutral currents -- $[ud]\leftrightarrow \{ud\}$.  In the isospin symmetry limit, the coupling strengths are identical in magnitude.  Using Eqs.\,\eqref{EqWeakq}, \eqref{EqSeagull} in Fig.\,\ref{FigWZqq}, one obtains
\begin{align}
\Gamma_{5\mu,\beta}^{sa}(p_d,k_d) & = \bigg[\kappa^{sa}_{\rm ax} m_N
\delta_{\mu\beta} - 2 \kappa^{sa}_{\rm ps} M_q^E \frac{Q_\mu Q_\beta}{Q^2+m_\pi^2}  \bigg]
d(\tau^{sa})\,.
\end{align}
Here $M_q^E=0.33\,$GeV is the Euclidean constituent quark mass defined by the dressed-quark propagator \cite[Eq.\,(A8)]{Chen:2019fzn}; $\tau^{sa}=Q^2/[4 m_{0^+} m_{1^+}]$; the computed $Q^2=0$ values of the couplings are
\begin{equation}
\label{EqKappaV1}
\kappa_{\rm ax}^{sa} = 0.75\,,\;
\kappa_{\rm ps}^{sa} = \kappa_{\rm ax}^{sa} \, m_N/[2 M_q^E] = 1.34\,,
\end{equation}
where the second equation expresses a Ward-Green-Takahashi identity, which we verified numerically;
and, emulating the electromagnetic current construction, $d(x)=1/(1+x)$ is introduced to express diquark compositeness via form factor suppression on $Q^2>0$.  Notably, using dipole suppression instead, \emph{i.e}.\ $d(x)^2$, no prediction in any image drawn herein changes by more than the line width because, in all cases, diagram (1) in Fig.\,\ref{FigCurrent} is both dominant and hard, whereas all weak-boson--diquark interactions are soft and subdominant.  (This is discussed further below.)
Note, too: $\Gamma_{5\mu,\beta}^{as} = -\Gamma_{5\mu,\beta}^{sa}$.

There are also transitions between axial-vector diquarks: charged currents -- $\{dd\}\leftrightarrow\{ud\}$, $\{ud\}\leftrightarrow\{uu\}$; and neutral currents -- $\{dd\}\leftrightarrow\{dd\}$, $\{uu\}\leftrightarrow\{uu\}$.  Once again, in the isospin symmetry limit, the coupling strengths are identical in magnitude.  Using Eqs.\,\eqref{EqWeakq}, \eqref{EqSeagull} in Fig.\,\ref{FigWZqq}, one obtains:
\begin{align}
i\Gamma_{5\mu,\alpha\beta}^{aa}&(p_d,k_d)  =
\bigg[ \kappa_{\rm ax}^{aa}\tfrac{1}{2}\varepsilon_{\mu\alpha\beta\rho}(p_d+k_d)_\rho
\nonumber \\
& \quad -\kappa_{\rm ps}^{aa} \frac{M_q^E}{m_N}  \frac{Q_\mu Q_\rho}{Q^2+m_\pi^2}
\varepsilon_{\alpha\beta\rho\sigma}
(p_d+k_d)_\sigma
\bigg]d(\tau^{aa})\,,
\end{align}
where $\tau^{aa} = Q^2/[4 m_{1^+}^2]$ and
\begin{equation}
\kappa_{\rm ax}^{aa} = 0.73\,,\;
\kappa_{\rm ps}^{aa} = \kappa_{\rm ax}^{aa} \, m_N/[2 M_q^E] = 1.31\,.
\end{equation}

It is worth highlighting that once the two diquark masses, Eq.\,\eqref{Eqqqmasses}, are fixed by requiring given values of the nucleon and $\Delta$-baryon masses, there are no free parameters in this study.  Both the Faddeev amplitude and the weak current are completely determined, with construction of the latter being constrained by the axial-vector Ward-Green-Takahashi identity.  Naturally, assumptions have been made in formulating the Faddeev kernel and weak current; but since we have closely followed the successful kindred treatment of electromagnetic processes, we judge those assumptions to be sound.

\begin{table}[t]
\caption{\label{TableI}
Selected $Q^2\simeq 0$ properties of the nucleon's axial-vector form factor, $G_A$, compared with RL-truncation three-body Faddeev equation results \cite{Eichmann:2011pv}, empirical determinations \cite{Zyla:2020, Liesenfeld:1999mv, Meyer:2016oeg, AguilarArevalo:2010zc}, and lattice-QCD (lQCD) computations \cite{Alexandrou:2017hac, Jang:2019vkm, Bali:2019yiy}.
Ref.\,\cite{Bali:2019yiy} reports results from two different parametrisations of their simulation outputs.
``--'' in any position indicates no information available in the cited source for the associated quantity.
The listed uncertainty in our predictions reflects the impact of $\pm 5$\% variations in the diquark masses in Eq.\,\eqref{Eqqqmasses}.
For comparison, the proton's measured electromagnetic radius yields $m_N \langle r_E^2\rangle^{1/2} = 4.00$ \cite{Zyla:2020}.}
\begin{center}
\begin{tabular*}
{\hsize}
{
l@{\extracolsep{0ptplus1fil}}
|l@{\extracolsep{0ptplus1fil}}
l@{\extracolsep{0ptplus1fil}}
l@{\extracolsep{0ptplus1fil}}}\hline
 & $g_A$ & $m_N\langle r_A^2\rangle^{1/2}$ & $m_A/m_N$ \\\hline
 Herein & 1.25(03) & 3.25(04) & 1.23(03) \\\hline
Faddeev$_3$\,\cite{Eichmann:2011pv} & 0.99(02) & 2.63(06) & 1.32(03) \\\hline
Exp\,\cite{Zyla:2020} & 1.2756(13) & -- & -- \\
Exp\,\cite{Liesenfeld:1999mv} & -- & 3.02(11) & 1.15(04) \\
Exp\,\cite{Meyer:2016oeg} & -- & 3.23(72) & 1.15(08) \\
Exp\,\cite{AguilarArevalo:2010zc} & -- & 2.41(31) & 1.44(18) \\\hline
lQCD\,\cite{Alexandrou:2017hac} & 1.21(3)(2) & 2.45(08)(03) & 1.41(04)(02) \\
lQCD\,\cite{Jang:2019vkm} & 1.30(6) & 3.57(30) & 0.97(16) \\
lQCD$_d$\,\cite{Bali:2019yiy} & 1.23(3) & 2.48(15) & 1.39(09) \\
lQCD$_z$\,\cite{Bali:2019yiy} & 1.30(9) & 3.19(30) & 1.09(11) \\\hline
\end{tabular*}
\end{center}
\end{table}

\section{Axial Form Factor}
\label{SecGA}
Using the elements described in Sec.\,\ref{SecNAC}, one can evaluate the sum of diagrams in Fig.\,\ref{FigCurrent} and, employing an appropriate spinor projection matrix, obtain the nucleon's axial form factor, $G_A(Q^2)$.  In Table~\ref{TableI}, we list predictions for selected $G_A(Q^2)$ characteristics in comparison with other determinations, with
\begin{equation}
\langle r_A^2 \rangle = - \frac{6}{G_A(0)} \left. \frac{d}{dQ^2} G_A(Q^2) \right|_{Q^2=0}.
\end{equation}
As signalled above, the listed uncertainties in our values express the impact of varying the diquark masses in Eq.\,\eqref{Eqqqmasses}.  The results obtained from the independent variations are combined with weight determined by the relative strength of scalar and axial-vector diquark contributions to $g_A=G_A(0)$, \emph{i.e}.\ approximately 4:1.  Notably, scalar and axial-vector diquark variations interfere destructively, \emph{e.g}.\ reducing $m_{[ud]}$ increases $g_A$, whereas $g_A$ decreases with the same change in the axial-vector mass.

\begin{figure}[t]
\leftline{\hspace*{0.5em}{\large{\textsf{A}}}}
\vspace*{-5ex}
\includegraphics[clip, width=0.42\textwidth]{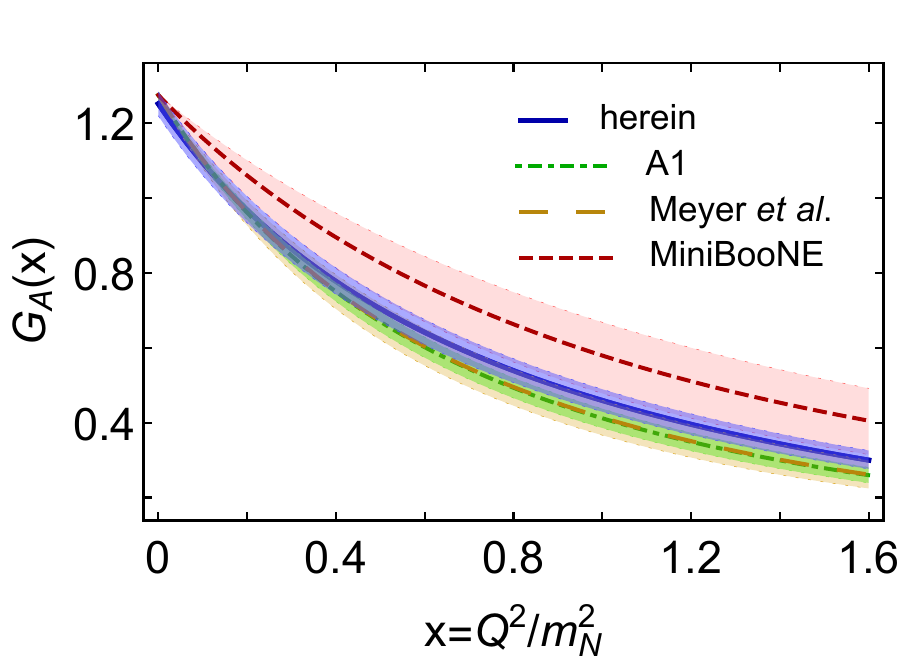}
\vspace*{1ex}

\leftline{\hspace*{0.5em}{\large{\textsf{B}}}}
\vspace*{-5ex}
\includegraphics[clip, width=0.42\textwidth]{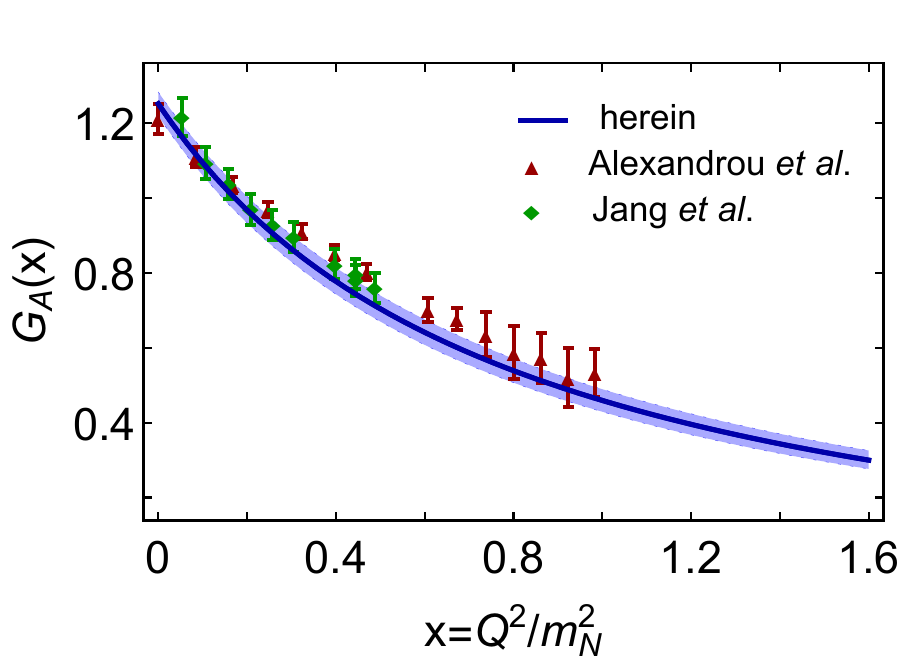}
\caption{\label{FigGAx}
\emph{Upper panel}\,--\,{\sf A}.  Result for $G_A(x)$ calculated herein -- blue curve within lighter blue model-uncertainty band, accompanied by empirical results report elsewhere: Ref.\,\cite{Liesenfeld:1999mv} -- dot-dashed green curve within like-coloured band; Ref.\,\cite{Meyer:2016oeg} -- long-dashed gold curve within like-coloured band; and Ref.\,\cite{AguilarArevalo:2010zc} -- dashed red curve within like-coloured band.
\emph{Lower panel}\,--\,{\sf B}.  Result for $G_A(x)$ calculated herein -- blue curve within lighter blue model-uncertainty band, compared with lQCD results: Ref.\,\cite{Alexandrou:2017hac} -- red up-triangles; and Ref.\,\cite{Jang:2019vkm} -- green diamonds.
These comparisons may be quantified by reporting the mean-$\chi^2$ values with respect to our central result, Eq.\,\eqref{EqGAx}, which are $4.36$ and $0.79$, respectively: our prediction and that in Ref.\,\cite{Jang:2019vkm} are in accord.  (With reference to Ref.\,\cite{Bali:2019yiy}, the mean-$\chi^2$ values are $3.18$ [dipole], $2.21$ [$z$ expansion].)
}
\end{figure}

Our prediction for the momentum dependence of $G_A$ is drawn in Fig.\,\ref{FigGAx}.  On the domain depicted, an accurate interpolation of the central result is provided by
\begin{equation}
\label{EqGAx}
G_A(x) = \frac{1.25 - 0.22 x}{1+1.24 x +0.0052 x^2}
\end{equation}
or, almost equally well, by a dipole characterised by the mass $M_A = 1.23\,m_N$.  Computed as explained above, the lighter blue band surrounding our result expresses the impact of $\pm 5$\% variations in the diquark masses that define the Faddeev kernel. 

\begin{table}[t]
\caption{\label{tablegr}
Referring to Fig.\,\ref{FigCurrent}, separation of $G_A(0)$ and $G_P(0)$ into contributions from various diagrams, listed as a fraction of the total $Q^2=0$ value.
Diagram (1): $\langle J \rangle^{S}_{\rm q}$ -- weak-boson strikes dressed-quark with scalar diquark spectator; and $\langle J \rangle^{A}_{\rm q}$ -- weak-boson strikes dressed-quark with axial-vector diquark spectator.
Diagram (2): $\langle J \rangle^{AA}_{\rm qq}$ -- weak-boson interacts strikes axial-vector diquark with dressed-quark spectator.
Diagram (3): $\langle J \rangle^{SA+AS}_{\rm dq}$ -- weak-boson mediates transition between scalar and axial-vector diquarks, with dressed-quark spectator.
Diagram (4): $\langle J \rangle_{\rm ex}$ -- weak-boson strikes dressed-quark ``in-flight'' between one diquark correlation and another.
Diagrams (5) and (6): $\langle J \rangle_{\rm sg}$ -- weak-boson couples inside the diquark correlation amplitude.
The listed uncertainty in these results reflects the impact of $\pm 5$\% variations in the diquark masses in Eq.\,\eqref{Eqqqmasses}, \emph{e.g}.\ $0.71_{1_\mp} \Rightarrow 0.71 \mp 0.01$. }
\begin{center}
\begin{tabular*}
{\hsize}
{
l@{\extracolsep{0ptplus1fil}}
|l@{\extracolsep{0ptplus1fil}}
l@{\extracolsep{0ptplus1fil}}
l@{\extracolsep{0ptplus1fil}}
l@{\extracolsep{0ptplus1fil}}
l@{\extracolsep{0ptplus1fil}}
l@{\extracolsep{0ptplus1fil}}}\hline
 & $\langle J \rangle^{S}_{\rm q}$  & $\langle J \rangle^{A}_{\rm q}$ &$\langle J \rangle^{AA}_{\rm qq}$ & $\langle J \rangle^{SA+AS}_{\rm qq}$ & $\langle J \rangle_{\rm ex}$ & $\langle J \rangle_{\rm sg}$ \\\hline
 $G_A(0)\ $ & $0.71_{4_\mp}$ & $0.064_{2_\pm} $ & $0.025_{5_\pm}$ & $0.13_{0_\mp}$ & $0.072_{32_\pm}$ & $\phantom{-}0$\\
 $G_P(0)\ $ & $0.74_{4_\mp}$ & $0.070_{5_\pm} $ & $0.025_{5_\pm}$ & $0.13_{0_\mp}$ & $0.22_{4_\pm}$ & $-0.19_{1_\mp}$\\  \hline
\end{tabular*}
\end{center}
\end{table}
In Table~\ref{tablegr}, referring to Fig.\,\ref{FigCurrent}, we list the relative strengths of each diagram's contribution to the nucleon's axial charge.  Diagram (1), with the weak-boson striking the dressed-quark in association with a spectator scalar diquark, is overwhelmingly dominant.  On the other hand, as anticipated following Eqs.\,\eqref{EqSeagull}, diagrams (5) and (6) are identically zero in this case.
%

\smallskip

\hspace*{-\parindent}\emph{Flavour Separation of $g_A$}\,---\,%
With some reflection, it becomes apparent that the fractions in the top row of Table~\ref{tablegr} can readily be translated into a flavour decomposition of the nucleon's axial charge; hence, the strength of $u$ and $d$ quark contributions to the nucleon light-front helicity.  For $u$-quarks and $d$-quarks, we find that the helicity parallel-antiparallel differences are $g_A^u=0.86 g_A$ and $g_A^d=-0.14 g_A$.  Hence, the nucleon's light-front helicity is overwhelmingly invested in the $u$-quark: $g_A^d/g_A^u = -0.16(2)$.  This result is consistent with the estimates made in Refs.\,\cite{Roberts:2013mja} using other means of analysing the nucleon's Faddeev amplitude.
It is notable that $g_A^d/g_A^u = -0.25$ in nonrelativistic quark models with uncorrelated wave functions \cite{He:1994gz}.  The difference between that value and our prediction highlights the impact of strong diquark correlations in the nucleon wave function: with high probability, the nucleon's $d$ quark is sequestered within a $[ud]$ diquark, which does not participate in weak interactions.

In contrast, contemporary lQCD analyses report values for this ratio that are larger in magnitude than the quark model result, locating a greater portion of the nucleon's axial charge with the $d$-quark:
$g_A^d/g_A^u = -0.40(2)$ \cite{Bhattacharya:2016zcn};
$g_A^d/g_A^u = -0.58(3)$ \cite{Alexandrou:2019brg}.
Notably, whilst $g_A$ is a conserved charge, invariant under QCD evolution \cite{Dokshitzer:1977sg, Gribov:1972, Lipatov:1974qm, Altarelli:1977}, the separation into component contributions from different quark flavours is not: $g_A^u$, $g_A^d$, etc.\ evolve individually in a manner which ensures the linear combination that defines $g_A$ is constant.  It follows that $g_A^d/g_A^u$ changes under QCD evolution.
This effect can potentially reconcile our flavour separation results with the lQCD predictions in Ref.\,\cite{Bhattacharya:2016zcn}: our Faddeev wave function is defined at the hadronic scale, $\zeta_H \approx 0.33\,$GeV \cite{Cui:2020dlm, Cui:2020piK}, whereas the lQCD values are renormalised at $\zeta \approx 2\,$GeV.
Adopting this perspective, the lQCD results in \cite{Alexandrou:2019brg} connect more closely with the uncorrelated quark model wave function.

The lQCD ratios are also interesting for another reason.  In the nonrelativistic quark model, $g_A^d/g_A^u = g_T^d/g_T^u$, where $g_T^{d,u}$ are the proton's tensor charges; and the ratio $g_T^d/g_T^u$ is renormalisation scale invariant.   Analogous lQCD ratios in this case are: $g_T^d/g_T^u = -0.25(2)$ \cite{Bhattacharya:2016zcn, Gupta:2018lvp}; $g_T^d/g_T^u =-0.29(3)$ \cite{Alexandrou:2019brg}.  The former value is consistent with the quark model result, whereas the latter is larger in magnitude.

The tensor charge ratio has not been computed in the framework employed herein; but that calculation was completed using the three-body RL Faddeev equation approach, with the result $g_T^d/g_T^u = -0.24(1)$ \cite{Wang:2018kto}.  As remarked above, the RL truncation underestimates the strengths of correlations in baryon wave functions.  Hence, we expect the framework used herein to yield a value for the ratio that is somewhat smaller in magnitude.  This conjecture will be tested in future.

The observations in this subsection bear on the proton spin puzzle, which has a thirty-year history \cite{Ashman:1987hv}.  They will subsequently be expounded upon elsewhere \cite{Chen:2020:progressSpin}.

\begin{figure}[t]
\leftline{\hspace*{0.5em}{\large{\textsf{A}}}}
\vspace*{-5ex}
\includegraphics[clip, width=0.42\textwidth]{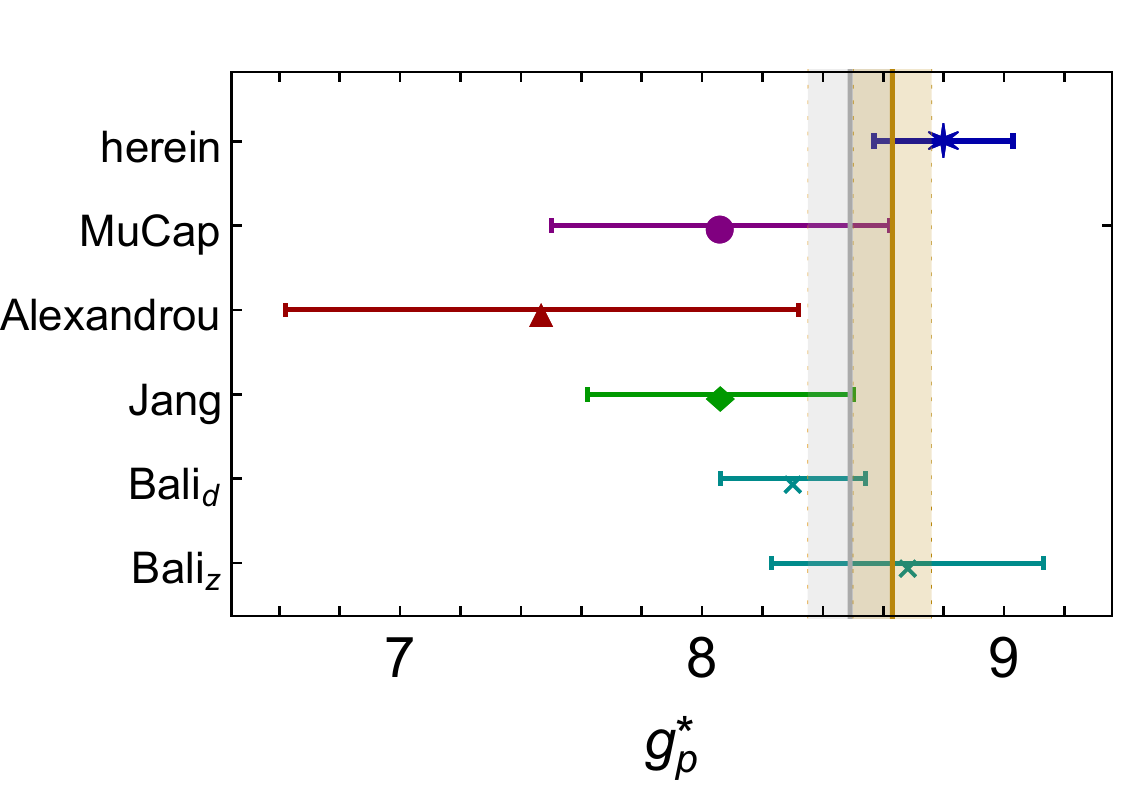}
\vspace*{1ex}

\leftline{\hspace*{0.5em}{\large{\textsf{B}}}}
\vspace*{-5ex}
\includegraphics[clip, width=0.42\textwidth]{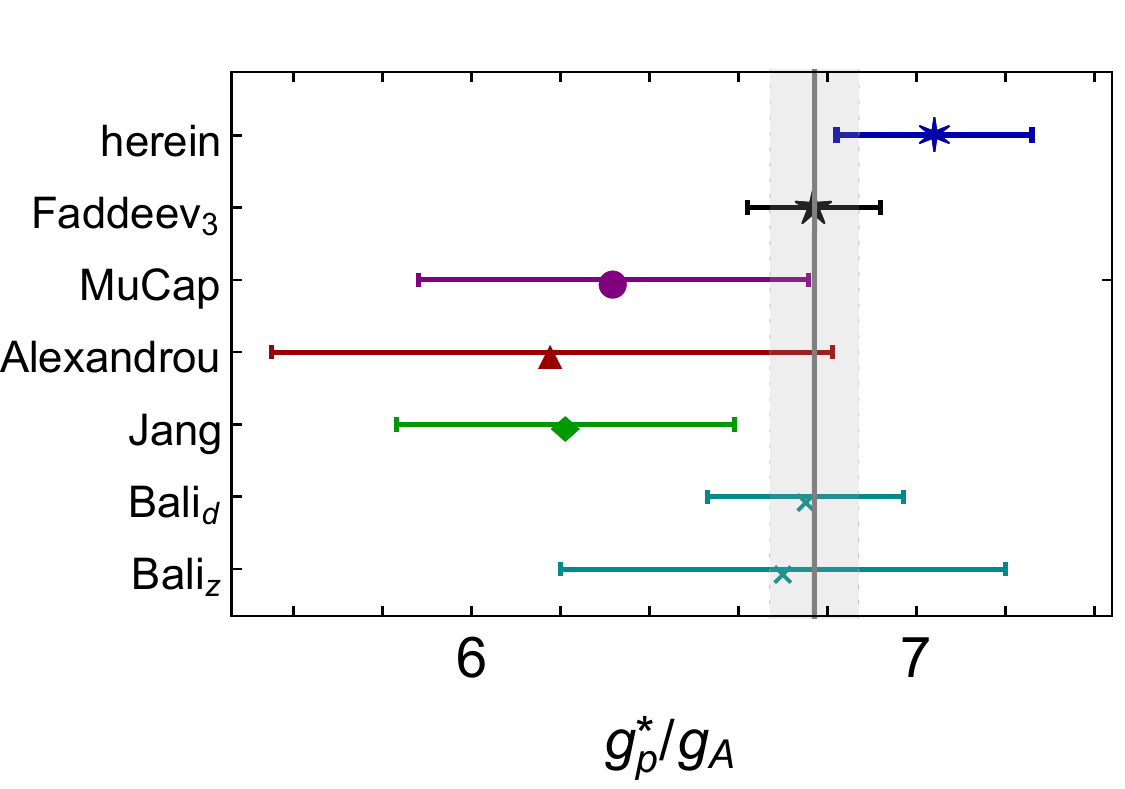}
\caption{\label{Figgpasterisk}
\emph{Upper panel}\,--\,{\sf A}.
Comparison of our prediction for $g_p^\ast$ (blue asterisk) with
an empirical value \cite{Andreev:2012fj} (purple circle)
and a collection of lQCD results: red triangle \cite{Alexandrou:2017hac}, green diamond \cite{Jang:2019vkm}, and cyan crosses \cite{Bali:2019yiy}.
An uncertainty-weighted average of the theory results is depicted by the grey line and associated band: $\bar g_p^\ast = 8.49(14)$.
\emph{Lower panel}\,--\,{\sf B}.  Analogous comparison for the ratio $g_P^\ast/g_A$, including the value computed from results in Ref.\,\cite{Eichmann:2011pv} (black star).  Again, the grey line and associated band depicts an uncertainty-weighted average of the theory results: $\overline{g_p^\ast/g_A} = 6.77(10)$.  Using the empirical result for $g_A=1.2756(13)$, this average value corresponds to $g_p^\ast=8.63(13)$, which is drawn as the gold line and associated band in the upper panel.
}
\end{figure}

\section{Induced Pseudoscalar Form Factor}
\label{SecGP}
$G_P(Q^2)$ in Eq.\,\eqref{EqJ5A} can be obtained following the procedure described in Sec.\,\ref{SecGA} simply by changing the spinor projection matrix.  The value at $Q^2=0.88m_\mu^2$, where $m_\mu$ is the mass of the $\mu$-lepton, defines the nucleon's induced pseudoscalar charge:
\begin{equation}
g_p^\ast = \frac{m_\mu}{2 m_N} G_P(Q^2 = 0.88 m_\mu^2),
\end{equation}
which can be measured in $\mu p $ capture reactions \cite{Gorringe:2002xx}, \emph{e.g}.\ $\mu\,+\,p\,\to\,\nu_\mu\,+\,n$.  We obtain $g_p^\ast = 8.80(23)$, a value that is consistent with both measurements and computations using other methods \cite[Fig.\,16]{Bali:2019yiy}: $8\lesssim g_p^\ast \lesssim 9$.
This is illustrated in Fig.\,\ref{Figgpasterisk}A, which depicts our prediction along with the value reported from a precision $\mu$-capture experiment \cite{Andreev:2012fj} and a collection of lQCD results \cite{Alexandrou:2017hac, Jang:2019vkm, Bali:2019yiy}.

It is worth remarking here that after rescaling to correct for the $g_A$ underestimate -- see Table~\ref{TableI}, the Ref.\,\cite{Eichmann:2011pv} RL-truncation three-body Faddeev equation analysis yields $g_p^\ast = 8.59(13)$.  The agreement between this value and the uncertainty-weighted average in Fig.\,\ref{Figgpasterisk}A suggests that some systematic theory uncertainties cancel in the ratio $g_P^\ast/g_A$.  So, in Fig.\,\ref{Figgpasterisk}B, we depict our prediction for $g_P^\ast/g_A = 7.04(22)$ along with that calculated using inputs from Ref.\,\cite{Eichmann:2011pv}, \emph{viz}.\ $g_P^\ast/g_A = 6.77(15)$, the value computed from a $\mu$-capture experiment \cite{Andreev:2012fj} and the empirical value of $g_A$, plus the ratios obtained from lQCD results \cite{Alexandrou:2017hac, Jang:2019vkm, Bali:2019yiy}.  Evidently, working with this ratio, there is some improvement in agreement between theory predictions.

\begin{figure}[t]
\includegraphics[clip, width=0.42\textwidth]{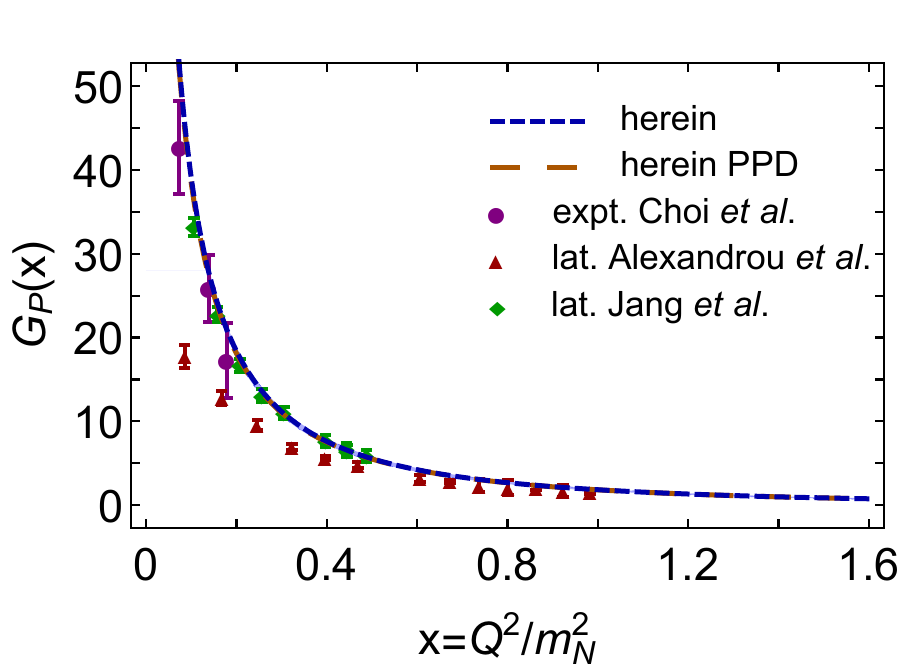}
\caption{\label{FigGPx}
Result for $G_P(x)$ calculated herein -- short-dashed blue curve.  (On the scale used in this figure, the associated lighter blue model-uncertainty band is practically invisible.)
The long-dashed orange curve is obtained using the pion pole dominance (PPD) approximation, Eq.\,\eqref{EqPPD}, with $G_A(x)$ in Eq.\,\eqref{EqGAx}.  Evidently, short-dashed-blue and long-dashed-orange curves agree within the line width.
For comparison, we also plot the data (purple circles) from Ref.\,\cite{Choi:1993vt} and lQCD results: Ref.\,\cite{Alexandrou:2017hac} -- red up-triangles; and Ref.\,\cite{Jang:2019vkm} -- green diamonds.
Considering the empirical data, the comparison may be quantified by reporting the mean-$\chi^2$ value, which is $1.41$.
In connection with the lQCD results, the mean-$\chi^2$ values are $61.9$ and $1.84$, respectively: our prediction agrees well with that in Ref.\,\cite{Jang:2019vkm}.
 (With reference to Ref.\,\cite{Bali:2019yiy}, the mean-$\chi^2$ values are $15.1$ [dipole], $3.84$ [$z$ expansion].)
}
\end{figure}

Our prediction for the momentum dependence of $G_P$ is drawn in Fig.\,\ref{FigGPx}.
On the domain depicted, an accurate interpolation of the central result is provided by
\begin{equation}
G_P(x) = \frac{40.1 -2.11 x -1.84 x^2}{0.127+ 7.75 x + 11.9 x^2}\,.
\end{equation}
The model-uncertainty band surrounding this curve is practically invisible on the ordinate scale necessary to draw the figure.

Returning to Eq.\,\eqref{WGTIexplicit}, it is evident that in the neighbourhood of the chiral limit, the right-hand-side is O$(m_\pi^2)$ and can be neglected; hence, it is a good approximation to write
\begin{equation}
\label{EqPPD}
G_P(x) \approx \frac{4}{x+m_\pi^2/m_N^2} G_A(x)\,.
\end{equation}
Treating this as an identity, one has the so-called pion pole dominance (PPD) prediction for $G_P$.  Looking at Fig.\,\ref{FigGPx}, it is evidently a good approximation, as observed empirically in Ref.\,\cite{Choi:1993vt}.  One can also illustrate the accuracy of PPD by seeking a least-squares fit to $G_P(x)$ using Eq.\,\eqref{EqPPD} and a dipole \emph{Ansatz} for $G_A(x)$: on $2 m_\pi^2/m_N^2<x<1.8$, which excises a small neighbourhood of the true pion pole, one finds $G_A(x) = 1.25/(1+x/1.24^2)^2$, \emph{i.e}.\ one effectively reproduces Eq.\,\eqref{EqGAx}.

In Table~\ref{tablegr}, referring to Fig.\,\ref{FigCurrent}, we also list the relative strengths of each diagram's contribution to the nucleon's induced pseudoscalar charge.  Once again, diagram (1), with the weak-boson striking the dressed-quark in association with a spectator scalar diquark, is overwhelmingly dominant.  In this case, however, there is an active cancellation between the contributions from diagrams (4), (5) and (6).


\section{Summary and Perspective}
Using a Poincar\'e-covariant quark+diquark Faddeev equation treatment of the nucleon and a weak interaction current that ensures consistency with relevant Ward-Green-Takahashi identities, we delivered
predictions for the nucleon's axial and induced pseudoscalar form factors, $G_A$ and $G_P$, respectively.  In doing so, we unified these features of the nucleon with an array of other properties of baryons and their excitations, \emph{e.g}.\  spectra and structural properties, as exposed in elastic and transition form factors.

Relating to the milieu created by modern neutrino experiments, our results may be summarised succinctly as follows.
$G_A$ can reliably be represented by a dipole form factor, normalised by the axial charge $g_A=1.25(3)$ and characterised by a mass-scale $M_A = 1.23(3) m_N$, where $m_N$ is the nucleon mass; and on their common domain, the associated pointwise behaviour matches well with that determined in a recent lattice-QCD (lQCD) study \cite{Jang:2019vkm}.
Based on the analysis in Ref.\,\cite{Lovato:2020kba}, which assumes nucleon properties are not materially modified in nuclei, one may reasonably expect that our prediction for $G_A$ can provide a sound foundation for the calculation of the neutrino-nucleus and antineutrino-nucleus cross-sections that are relevant, \emph{e.g}.\ to modern accelerator neutrino experiments.

Regarding $G_P$: our predictions for the induced pseudoscalar charge, $g_p^\ast$, and the ratio $g_p^\ast/g_A$ are consistent with the values determined from the most recent $\mu$-capture experiment \cite{Andreev:2012fj}; and its momentum-dependence agrees with data obtained from low-energy pion electroproduction \cite{Choi:1993vt} and also the lQCD result in Ref.\,\cite{Jang:2019vkm}.  Furthermore, we find that the pion pole dominance \emph{Ansatz} provides a sound estimate of the directly computed result, \emph{viz}.\ using a ${\cal L}_1$ measure, the mean relative difference between the curves is 2.7\% on $0 \leq Q^2 \leq 1.6 m_N^2$.


It is worth highlighting that we find $g_A^d/g_A^u=-0.16(2)$ at the hadronic scale.  This is a significant suppression of the magnitude of the $d$-quark component relative to that found in nonrelativistic quark models.  The size reduction owes to the presence of strong diquark correlations in our nucleon wave function, with the calculated value reflecting the relative strength of scalar and axial-vector diquarks: the isoscalar--scalar correlations are dominant, but the isovector--axial-vector diquarks have a measurable influence.

Herein, we canvassed physics impacts of our treatment of the nucleon axial current.  Unification of this analysis with that of the nucleon pseudoscalar current, Eq.\,\eqref{EqJ5}; a detailed discussion of the partial conservation of the nucleon axial current and associated Goldberger-Treiman relations; and all technical details relating to current constructions in our quark+diquark approach, including the seagull terms, will be presented elsewhere \cite{Chen:2020:progress}.

A natural next step is to go beyond the quark+diquark approach and use the more fundamental three-quark Faddeev equation treatment of the baryon problem, revisiting the analysis in Ref.\,\cite{Eichmann:2011pv}.  In seeking and following a path toward improving the expressions of emergent hadronic mass in both the Faddeev kernel and interaction current, one could therewith provide continuum predictions for the nucleon's axial current form factors that posses a more rigorous connection to QCD's Schwinger functions.

\smallskip

\noindent\textbf{Acknowledgments}\,---\,
%
We are grateful to Y.-C.~Jang for providing us with the lattice results in Ref.\,\cite{Jang:2019vkm} and for constructive comments from Z.-F.~Cui, G.~Eichmann, M.~Engelhardt, A.~Lovato, U.~Mosel, M.~Oettel and N.~Santowsky.
Work supported by:
%
%
DFG grant FI 970/11-1;
Jiangsu Province \emph{Hundred Talents Plan for Professionals};
Chinese Ministry of Science and Technology \emph{International Expert Involvement Programme};
Ministerio Espa\~nol de Ciencia e Innovaci\'on, grant no. PID2019-107844GB-C22;
and Junta de Andaluc\'ia, contract nos.\ P18-FRJ-1132 and Operativo FEDER Andaluc\'ia 2014-2020 UHU-1264517.
%





\begin{thebibliography}{10}
\expandafter\ifx\csname url\endcsname\relax
  \def\url#1{\texttt{#1}}\fi
\expandafter\ifx\csname urlprefix\endcsname\relax\def\urlprefix{URL }\fi
\expandafter\ifx\csname href\endcsname\relax
  \def\href#1#2{#2} \def\path#1{#1}\fi

\bibitem{RevModPhys.88.030501}
T.~Kajita, Nobel lecture: Discovery of atmospheric neutrino oscillations, Rev.
  Mod. Phys. 88 (2016) 030501.

\bibitem{RevModPhys.88.030502}
A.~B. McDonald, Nobel lecture: The sudbury neutrino observatory: Observation of
  flavor change for solar neutrinos, Rev. Mod. Phys. 88 (2016) 030502.

\bibitem{Gouvea:2016shl}
A.~de~Gouv\^ea, {Neutrino Mass Models}, Ann. Rev. Nucl. Part. Sci. 66 (2016)
  197--217.

\bibitem{Zyla:2020}
P.~A. Zyla, et~al., Review of particle properties, Prog. Theor. Exp. Phys.
  083C01.

\bibitem{Mosel:2016cwa}
U.~Mosel, {Neutrino Interactions with Nucleons and Nuclei: Importance for
  Long-Baseline Experiments}, Ann. Rev. Nucl. Part. Sci. 66 (2016) 171--195.

\bibitem{Alvarez-Ruso:2017oui}
L.~Alvarez-Ruso, et~al., {NuSTEC White Paper: Status and challenges of
  neutrino\textendash{}nucleus scattering}, Prog. Part. Nucl. Phys. 100 (2018)
  1--68.

\bibitem{Hill:2017wgb}
R.~J. Hill, P.~Kammel, W.~J. Marciano, A.~Sirlin, {Nucleon Axial Radius and
  Muonic Hydrogen \textemdash{} A New Analysis and Review}, Rept. Prog. Phys.
  81 (2018) 096301.

\bibitem{Gysbers:2019uyb}
P.~Gysbers, et~al., {Discrepancy between experimental and theoretical
  $\beta$-decay rates resolved from first principles}, Nature Phys. 15~(5)
  (2019) 428--431.

\bibitem{Lovato:2020kba}
A.~Lovato, J.~Carlson, S.~Gandolfi, N.~Rocco, R.~Schiavilla, {Ab initio study
  of $\boldsymbol{(\nu_\ell,\ell^-)}$ and
  $\boldsymbol{(\overline{\nu}_\ell,\ell^+)}$ inclusive scattering in $^{12}$C:
  confronting the MiniBooNE and T2K CCQE data}, Phys. Rev. X 10 (2020) 031068.

\bibitem{King:2020wmp}
G.~B. King, L.~Andreoli, S.~Pastore, M.~Piarulli, R.~Schiavilla, R.~B. Wiringa,
  J.~Carlson, S.~Gandolfi, {Chiral Effective Field Theory Calculations of Weak
  Transitions in Light Nuclei}, Phys. Rev. C 102~(2) (2020) 025501.

\bibitem{Deur:2018roz}
A.~Deur, S.~J. Brodsky, G.~F. De~T\'eramond, {The Spin Structure of the
  Nucleon}, Rept. Prog. Phys. 82~(076201).

\bibitem{Baker:1981su}
N.~J. Baker, A.~M. Cnops, P.~L. Connolly, S.~A. Kahn, H.~G. Kirk, M.~J.
  Murtagh, R.~B. Palmer, N.~P. Samios, M.~Tanaka, {Quasielastic Neutrino
  Scattering: A Measurement of the Weak Nucleon Axial Vector Form-Factor},
  Phys. Rev. D 23 (1981) 2499--2505.

\bibitem{Miller:1982qi}
K.~L. Miller, et~al., {Study of the reaction $\nu d \to \mu^- p p_s$}, Phys.
  Rev. D 26 (1982) 537--542.

\bibitem{Kitagaki:1983px}
T.~Kitagaki, et~al., {High-energy quasielastic $\nu_\mu n \to \mu^- p$
  scattering in deuterium}, Phys. Rev. D 28 (1983) 436--442.

\bibitem{Ahrens:1986xe}
L.~A. Ahrens, et~al., {Measurement of neutrino-proton and antineutrino-proton
  elastic scattering}, Phys. Rev. D 35 (1987) 785.

\bibitem{Liesenfeld:1999mv}
A.~Liesenfeld, et~al., {A Measurement of the axial form-factor of the nucleon
  by the $p(e,e^\prime \pi^+)n$ reaction at $W = 1125\,$MeV}, Phys. Lett. B 468
  (1999) 20.

\bibitem{Meyer:2016oeg}
A.~S. Meyer, M.~Betancourt, R.~Gran, R.~J. Hill, {Deuterium target data for
  precision neutrino-nucleus cross sections}, Phys. Rev. D 93 (2016) 113015.

\bibitem{Gran:2006jn}
R.~Gran, et~al., {Measurement of the quasi-elastic axial vector mass in
  neutrino-oxygen interactions}, Phys. Rev. D 74 (2006) 052002.

\bibitem{Dorman:2009zz}
M.~Dorman, {Preliminary results for CCQE scattering with the MINOS near
  detector}, AIP Conf. Proc. 1189 (2009) 133--138.

\bibitem{AguilarArevalo:2010zc}
A.~Aguilar-Arevalo, et~al., {First Measurement of the Muon Neutrino Charged
  Current Quasielastic Double Differential Cross Section}, Phys. Rev. D 81
  (2010) 092005.

\bibitem{Fields:2013zhk}
L.~Fields, et~al., {Measurement of Muon Antineutrino Quasielastic Scattering on
  a Hydrocarbon Target at $E_\nu \sim 3.5$ GeV}, Phys. Rev. Lett. 111 (2013)
  022501.

\bibitem{Fiorentini:2013ezn}
G.~Fiorentini, et~al., {Measurement of Muon Neutrino Quasielastic Scattering on
  a Hydrocarbon Target at $E_\nu \sim 3.5$ GeV}, Phys. Rev. Lett. 111 (2013)
  022502.

\bibitem{Aubert:1983xm}
J.~Aubert, et~al., {The ratio of the nucleon structure functions $F_2^N$ for
  iron and deuterium}, Phys. Lett. B 123 (1983) 275.

\bibitem{Weinstein:2010rt}
L.~B. Weinstein, E.~Piasetzky, D.~W. Higinbotham, J.~Gomez, O.~Hen, et~al.,
  {Short Range Correlations and the EMC Effect}, Phys. Rev. Lett. 106 (2011)
  052301.

\bibitem{Lynn:2019rdt}
J.~E. Lynn, I.~Tews, S.~Gandolfi, A.~Lovato, {Quantum Monte Carlo Methods in
  Nuclear Physics: Recent Advances}, Ann. Rev. Nucl. Part. Sci. 69 (2019)
  279--305.

\bibitem{DeSanctis:1998ck}
M.~De~Sanctis, E.~Santopinto, M.~M. Giannini, {A Relativistic study of the
  nucleon form-factors}, Eur. Phys. J. A 1 (1998) 187--192.

\bibitem{DeSanctis:2011zz}
M.~De~Sanctis, J.~Ferretti, E.~Santopinto, A.~Vassallo, {Electromagnetic form
  factors in the relativistic interacting quark-diquark model of baryons},
  Phys. Rev. C 84 (2011) 055201.

\bibitem{Hagler:2009ni}
P.~H{\"a}gler, {Hadron structure from lattice quantum chromodynamics}, Phys.
  Rept. 490 (2010) 49--175.

\bibitem{Punjabi:2015bba}
V.~Punjabi, C.~F. Perdrisat, M.~K. Jones, E.~J. Brash, C.~E. Carlson, {The
  Structure of the Nucleon: Elastic Electromagnetic Form Factors}, Eur. Phys.
  J. A 51 (2015) 79.

\bibitem{DeGrand:1975cf}
T.~A. DeGrand, R.~Jaffe, K.~Johnson, J.~Kiskis, {Masses and Other Parameters of
  the Light Hadrons}, Phys. Rev. D 12 (1975) 2060.

\bibitem{Thomas:1981vc}
A.~W. Thomas, S.~Theberge, G.~A. Miller, {The Cloudy Bag Model of the Nucleon},
  Phys. Rev. D 24 (1981) 216.

\bibitem{Weise:1991}
W.~Weise, {Models of the nucleon}, in: {Proceedings of the 5th Swieca School in
  Nuclear Physics, Brazil. Bertulani, C. A. (ed.)}, 1991, pp. 175--244.

\bibitem{Alkofer:1994ph}
R.~Alkofer, H.~Reinhardt, H.~Weigel, {Baryons as chiral solitons in the
  Nambu-Jona-Lasinio model}, Phys. Rept. 265 (1996) 139--252.

\bibitem{Boffi:2001zb}
S.~Boffi, L.~Y. Glozman, W.~Klink, W.~Plessas, M.~Radici, R.~F. Wagenbrunn,
  {Covariant electroweak nucleon form-factors in a chiral constituent quark
  model}, Eur. Phys. J. A 14 (2002) 17--21.

\bibitem{Adamuscin:2007fk}
C.~Adamuscin, E.~Tomasi-Gustafsson, E.~Santopinto, R.~Bijker, {Two-component
  model for the axial form factor of the nucleon}, Phys. Rev. C 78 (2008)
  035201.

\bibitem{Kronfeld:2019nfb}
A.~S. Kronfeld, D.~G. Richards, W.~Detmold, R.~Gupta, H.-W. Lin, K.-F. Liu,
  A.~S. Meyer, R.~Sufian, S.~Syritsyn, {Lattice QCD and Neutrino-Nucleus
  Scattering}, Eur. Phys. J. A 55 (2019) 196.

\bibitem{Eichmann:2016yit}
G.~Eichmann, H.~Sanchis-Alepuz, R.~Williams, R.~Alkofer, C.~S. Fischer,
  {Baryons as relativistic three-quark bound states}, Prog. Part. Nucl. Phys.
  91 (2016) 1--100.

\bibitem{Papavassiliou:2017qlq}
J.~Papavassiliou, A.~C. Aguilar, D.~Binosi, C.~T. Figueiredo, {Mass generation
  in Yang-Mills theories}, EPJ Web Conf. 164 (2017) 03005.

\bibitem{Huber:2018ned}
M.~Q. Huber, {Nonperturbative properties of Yang-Mills theories}, Phys. Rept.
  879 (2020) 1 -- 92.

\bibitem{Fischer:2018sdj}
C.~S. Fischer, {QCD at finite temperature and chemical potential from
  Dyson--Schwinger equations}, Prog. Part. Nucl. Phys. 105 (2019) 1--60.

\bibitem{Roberts:2020hiw}
C.~D. Roberts, {Empirical Consequences of Emergent Mass}, Symmetry 12 (2020)
  1468.

\bibitem{Qin:2020rad}
S.-X. Qin, C.~D. Roberts, {Impressions of the Continuum Bound State Problem in
  QCD}, Chin. Phys. Lett. 37~(12) (2020) 121201.

\bibitem{Eichmann:2011pv}
G.~Eichmann, C.~S. Fischer, {Nucleon axial and pseudoscalar form factors from
  the covariant Faddeev equation}, Eur. Phys. J. A 48 (2012) 9.

\bibitem{Munczek:1994zz}
H.~J. Munczek, {Dynamical chiral symmetry breaking, Goldstone's theorem and the
  consistency of the Schwinger-Dyson and Bethe-Salpeter Equations}, Phys. Rev.
  D 52 (1995) 4736--4740.

\bibitem{Bender:1996bb}
A.~Bender, C.~D. Roberts, L.~von Smekal, {Goldstone Theorem and Diquark
  Confinement Beyond Rainbow- Ladder Approximation}, Phys. Lett. B 380 (1996)
  7--12.

\bibitem{Chang:2012cc}
L.~Chang, C.~D. Roberts, S.~M. Schmidt, {Dressed-quarks and the nucleon's axial
  charge}, Phys. Rev. C 87 (2013) 015203.

\bibitem{Segovia:2014aza}
J.~Segovia, I.~C. Cloet, C.~D. Roberts, S.~M. Schmidt, {Nucleon and $\Delta$
  elastic and transition form factors}, Few Body Syst. 55 (2014) 1185--1222.

\bibitem{Burkert:2017djo}
V.~D. Burkert, C.~D. Roberts, {Roper resonance: Toward a solution to the
  fifty-year puzzle}, Rev. Mod. Phys. 91 (2019) 011003.

\bibitem{Chen:2017pse}
C.~Chen, et~al., {Structure of the nucleon's low-lying excitations}, Phys. Rev.
  D 97 (2018) 034016.

\bibitem{Chen:2018nsg}
C.~Chen, et~al., {Nucleon-to-Roper electromagnetic transition form factors at
  large $Q^2$}, Phys. Rev. D 99 (2019) 034013.

\bibitem{Chen:2019fzn}
C.~Chen, G.~I. Krein, C.~D. Roberts, S.~M. Schmidt, J.~Segovia, {Spectrum and
  structure of octet and decuplet baryons and their positive-parity
  excitations}, Phys. Rev. D 100 (2019) 054009.

\bibitem{Lu:2019bjs}
Y.~Lu, C.~Chen, Z.-F. Cui, C.~D. Roberts, S.~M. Schmidt, J.~Segovia, H.~S.
  Zong, {Transition form factors: $\gamma^\ast + p \to \Delta(1232)$,
  $\Delta(1600)$}, Phys. Rev. D 100 (2019) 034001.

\bibitem{Cui:2020rmu}
Z.-F. Cui, C.~Chen, D.~Binosi, F.~de~Soto, C.~D. Roberts,
  J.~Rodr{\'{\i}}guez-Quintero, S.~M. Schmidt, J.~Segovia, {Nucleon elastic
  form factors at accessible large spacelike momenta}, Phys. Rev. D 102 (2020)
  014043.

\bibitem{Barabanov:2020jvn}
M.~Y. Barabanov, et~al., {Diquark Correlations in Hadron Physics: Origin,
  Impact and Evidence}, Prog. Part. Nucl. Phys. 116 (2021) 103835.

\bibitem{Cahill:1988dx}
R.~T. Cahill, C.~D. Roberts, J.~Praschifka, {Baryon structure and QCD},
  Austral. J. Phys. 42 (1989) 129--145.

\bibitem{Reinhardt:1989rw}
H.~Reinhardt, {Hadronization of Quark Flavor Dynamics}, Phys. Lett. B 244
  (1990) 316--326.

\bibitem{Efimov:1990uz}
G.~V. Efimov, M.~A. Ivanov, V.~E. Lyubovitskij, {Quark - diquark approximation
  of the three quark structure of baryons in the quark confinement model}, Z.
  Phys. C 47 (1990) 583--594.

\bibitem{Chang:2010hb}
L.~Chang, Y.-X. Liu, C.~D. Roberts, {Dressed-quark anomalous magnetic moments},
  Phys. Rev. Lett. 106 (2011) 072001.

\bibitem{Chang:2011ei}
L.~Chang, C.~D. Roberts, {Tracing masses of ground-state light-quark mesons},
  Phys. Rev. C 85 (2012) 052201(R).

\bibitem{Williams:2015cvx}
R.~Williams, C.~S. Fischer, W.~Heupel, {Light mesons in QCD and unquenching
  effects from the 3PI effective action}, Phys. Rev. D 93 (2016) 034026.

\bibitem{Ishii:1998tw}
N.~Ishii, {Meson exchange contributions to the nucleon mass in the Faddeev
  approach to the NJL model}, Phys. Lett. B 431 (1998) 1--7.

\bibitem{Hecht:2002ej}
M.~B. Hecht, C.~D. Roberts, M.~Oettel, A.~W. Thomas, S.~M. Schmidt, P.~C.
  Tandy, {Nucleon mass and pion loops}, Phys. Rev. C 65 (2002) 055204.

\bibitem{Sanchis-Alepuz:2014wea}
H.~Sanchis-Alepuz, C.~S. Fischer, S.~Kubrak, {Pion cloud effects on baryon
  masses}, Phys. Lett. B 733 (2014) 151--157.

\bibitem{Aznauryan:2012ba}
I.~Aznauryan, A.~Bashir, V.~Braun, S.~Brodsky, V.~Burkert, et~al., {Studies of
  Nucleon Resonance Structure in Exclusive Meson Electroproduction}, Int. J.
  Mod. Phys. E 22 (2013) 1330015.

\bibitem{Oettel:1999gc}
M.~Oettel, M.~Pichowsky, L.~von Smekal, {Current conservation in the covariant
  quark-diquark model of the nucleon}, Eur. Phys. J. A 8 (2000) 251--281.

\bibitem{Bloch:1999rm}
J.~C.~R. Bloch, C.~D. Roberts, S.~M. Schmidt, {Selected nucleon form-factors
  and a composite scalar diquark}, Phys. Rev. C 61 (2000) 065207.

\bibitem{Oettel:2000jj}
M.~Oettel, R.~Alkofer, L.~von Smekal, {Nucleon properties in the covariant
  quark diquark model}, Eur. Phys. J. A 8 (2000) 553--566.

\bibitem{Chen:2020:progress}
C.~Chen, C.~S. Fischer, C.~D. Roberts, J.~Segovia, {Form Factors of the Nucleon
  Axial and Pseudoscalar Currents -- \emph{in progress}}.

\bibitem{Alexandrou:2017hac}
C.~Alexandrou, M.~Constantinou, K.~Hadjiyiannakou, K.~Jansen, C.~Kallidonis,
  G.~Koutsou, A.~Vaquero Aviles-Casco, {Nucleon axial form factors using $N_f$
  = 2 twisted mass fermions with a physical value of the pion mass}, Phys. Rev.
  D 96 (2017) 054507.

\bibitem{Jang:2019vkm}
Y.-C. Jang, R.~Gupta, B.~Yoon, T.~Bhattacharya, {Axial Vector Form Factors from
  Lattice QCD that Satisfy the PCAC Relation}, Phys. Rev. Lett. 124 (2020)
  072002.

\bibitem{Bali:2019yiy}
G.~S. Bali, L.~Barca, S.~Collins, M.~Gruber, M.~L\"offler, A.~Sch\"afer,
  W.~S\"oldner, P.~Wein, S.~Weish\"aupl, T.~Wurm, {Nucleon axial structure from
  lattice QCD}, JHEP 05 (2020) 126.

\bibitem{Roberts:2013mja}
C.~D. Roberts, R.~J. Holt, S.~M. Schmidt, {Nucleon spin structure at very high
  $x$}, Phys. Lett. B 727 (2013) 249--254.

\bibitem{He:1994gz}
H.~He, X.~Ji, {The Nucleon's tensor charge}, Phys. Rev. D 52 (1995) 2960--2963.

\bibitem{Bhattacharya:2016zcn}
T.~Bhattacharya, V.~Cirigliano, S.~Cohen, R.~Gupta, H.-W. Lin, B.~Yoon, {Axial,
  Scalar and Tensor Charges of the Nucleon from 2+1+1-flavor Lattice QCD},
  Phys. Rev. D 94 (2016) 054508.

\bibitem{Alexandrou:2019brg}
C.~Alexandrou, S.~Bacchio, M.~Constantinou, J.~Finkenrath, K.~Hadjiyiannakou,
  K.~Jansen, G.~Koutsou, A.~Vaquero Aviles-Casco, {Nucleon axial, tensor, and
  scalar charges and $\sigma$-terms in lattice QCD}, Phys. Rev. D 102 (2020)
  054517.

\bibitem{Dokshitzer:1977sg}
Y.~L. Dokshitzer, Calculation of the structure functions for deep inelastic
  scattering and e+ e- annihilation by perturbation theory in quantum
  chromodynamics. ({\mbox {i}n {r}ussian}), Sov. Phys. JETP 46 (1977) 641--653.

\bibitem{Gribov:1972}
V.~N. Gribov, L.~N. Lipatov, Deep inelastic e p scattering in perturbation
  theory, Sov. J. Nucl. Phys. 15 (1972) 438--450.

\bibitem{Lipatov:1974qm}
L.~N. Lipatov, {The parton model and perturbation theory}, Sov. J. Nucl. Phys.
  20 (1975) 94--102.

\bibitem{Altarelli:1977}
G.~Altarelli, G.~Parisi, Asymptotic freedom in parton language, Nucl. Phys. B
  126 (1977) 298.

\bibitem{Cui:2020dlm}
Z.-F. Cui, M.~Ding, F.~Gao, K.~Raya, D.~Binosi, L.~Chang, C.~D. Roberts,
  J.~Rodr\'{\i}guez-Quintero, S.~M. Schmidt, {Higgs modulation of emergent mass
  as revealed in kaon and pion parton distributions}, Eur. Phys. J. A (Lett.)
  57~(1) (2021) 5.

\bibitem{Cui:2020piK}
Z.-F. Cui, M.~Ding, F.~Gao, K.~Raya, D.~Binosi, L.~Chang, C.~D. Roberts,
  J.~Rodr{\'{\i}}guez-Quintero, S.~M. Schmidt, {Kaon and pion parton
  distributions}, Eur. Phys. J. C 80 (2020) 1064.

\bibitem{Gupta:2018lvp}
R.~Gupta, B.~Yoon, T.~Bhattacharya, V.~Cirigliano, Y.-C. Jang, H.-W. Lin,
  {Flavor diagonal tensor charges of the nucleon from (2+1+1)-flavor lattice
  QCD}, Phys. Rev. D 98 (2018) 091501.

\bibitem{Wang:2018kto}
Q.-W. Wang, S.-X. Qin, C.~D. Roberts, S.~M. Schmidt, {Proton tensor charges
  from a Poincar{\'e}-covariant Faddeev equation}, Phys. Rev. D 98 (2018)
  054019.

\bibitem{Ashman:1987hv}
J.~Ashman, et~al., {A Measurement of the Spin Asymmetry and Determination of
  the Structure Function g(1) in Deep Inelastic Muon-Proton Scattering}, Phys.
  Lett. B 206 (1988) 364.

\bibitem{Chen:2020:progressSpin}
C.~Chen, et~al., {\emph{in progress}}.

\bibitem{Andreev:2012fj}
V.~Andreev, et~al., {Measurement of Muon Capture on the Proton to 1\% Precision
  and Determination of the Pseudoscalar Coupling $g_P$}, Phys. Rev. Lett. 110
  (2013) 012504.

\bibitem{Gorringe:2002xx}
T.~Gorringe, H.~W. Fearing, {Induced Pseudoscalar Coupling of the Proton Weak
  Interaction}, Rev. Mod. Phys. 76 (2004) 31--91.

\bibitem{Choi:1993vt}
S.~Choi, et~al., {Axial and pseudoscalar nucleon form-factors from low-energy
  pion electroproduction}, Phys. Rev. Lett. 71 (1993) 3927--3930.

\end{thebibliography}

\end{document}